\documentclass[preprint,showpacs,showkeys,preprintnumbers,prb]{revtex4}

\usepackage{graphicx}

\begin{document}

\title{Calculated electronic and magnetic properties of the half-metallic, transition metal based Heusler compounds.}

\author{Hem C. Kandpal, Gerhard H. Fecher, and Claudia Felser}
\email{felser@uni-mainz.de}
\affiliation{
Institut f\"ur Anorganische Chemie und Analytische Chemie, \\
Johannes Gutenberg - Universit\"at, D-55099 Mainz, Germany.}

\date{\today}

%%%%%%%%%%%%%%%%%%%%%%%%%%%%%%%%%%%%%%%%%%%%%%%%%%%%%%%%%%%%%%%%%%%%%%%
%%%%%%%%%%%%%%%Begin Abstract %%%%%%%%%%%%%%%%%%%%%%%%%%%%%%%%%%%%%%%%%
%%%%%%%%%%%%%%%%%%%%%%%%%%%%%%%%%%%%%%%%%%%%%%%%%%%%%%%%%%%%%%%%%%%%%%%

\begin{abstract}

In this work, results of {\it ab-initio} band structure calculations for $A_2BC$ Heusler compounds that have $A$ and $B$ sites occupied by transition metals and $C$ by a main group element are presented. This class of materials includes some interesting half-metallic and ferromagnetic properties. The calculations have been performed in order to understand the properties of the minority band gap and the peculiar magnetic behavior found in these materials. Among the interesting aspects of the electronic structure of the materials are the contributions from both $A$ and $B$ atoms to states near the Fermi energy and to the total magnetic moment. The magnitude of the total magnetic moment, which depends as well on the kind of $C$ atoms, shows a trend consistent with the Slater-Pauling type behavior in several classes of these compounds. The localized moment in these magnetic compounds resides at the $B$ site. Other than in the classical Cu$_2$-based Heusler compounds, the $A$ atoms in Co$_2$, Fe$_2$, and Mn$_2$ based compounds may contribute pronounced to the total magnetic moment.

%%%%%%%%%%%%%%%%%%%%%%%%%%%%%%%%%%%%%%%%%%%%%%%%%%%%%%%%%%%%%%%%%%%%%%%
%%%%%%%%%%%%%End Abstract %%%%%%%%%%%%%%%%%%%%%%%%%%%%%%%%%%%%%%%%%%%%%

\end{abstract}



\maketitle

%%%%%%%%%%%%%%%%%%%%%%Begin Section: Introduction%%%%%%%%%%%%%%%%%%%%%%

\section{Introduction}
The Heusler compounds \cite{Heu03} are ternary intermetallics with a 2:1:1 stoichiometry and the chemical formula $A_2BC$. They usually consist of two transition metals ($A$, $B$) and a main group element ($C$). They first attracted the interest to the magnetism community when Heusler had shown that the compound Cu$_x$Mn$_y$Al becomes ferromagnetic in the 211 form ($x$ = 2 and $y$ = 1), even none of its constituents is ferromagnetic by itself. However, it took three decades before their structure was explained to be an ordered compound \cite{BRo34,Heu34}.

The main interest during the first decades after their discovery was concentrated on Cu and Mn containing compounds. Co$_2$ based compounds were synthesized and investigated in the 1970s \cite{Web71}. K{\"u}bler {\it et al.} \cite{KWS83} recognized that the minority-spin state densities at the Fermi energy ($\epsilon_F$) nearly vanish for Co$_2$MnAl and Co$_2$MnSn. The authors concluded that this should lead to peculiar transport properties in these Heusler compounds because only the majority density contributes at $\epsilon_F$. At the same time, de Groot {\it et al.} \cite{GME83} proposed the concept of the so called half-metallic ferromagnets (HMF) that are materials predicted to exhibit 100~{\%} spin polarization at $\epsilon_F$. This exceptional property would make the HMF ideal candidates for spin injection devices to be used in spin electronics \cite{CVB02}.
  
The electronic structure plays an important role in determining the magnetic properties of Heusler compounds and, in particular, for predicting half-metallic ferromagnetism. Therefore, the band structure calculations must be performed very carefully. The first attempt to calculate the band structure of some Co$_2$-based compounds (Co$_2$MnSn, Co$_2$TiSi and Co$_2$TiAl) did not indicate half-metallic ferromagnetism \cite{IAK82}. These calculations displayed a minimum of the minority density of states (DOS) at $\epsilon_F$. At that time, the calculations were based on spherical potentials, and the exchange-correlation potential of the local spin density approximation (LSDA) was used in a rather simple form \cite{KSh65,HLu71,BHe72,VWN80}. The first clear indication of half-metallic ferromagnetism in Co$_2$-based Heusler compounds was reported by Ishida {\it et al.} \cite{IFK95, IKF95} for Co$_2$Mn$C$ and Ru$_2$Mn$C$ ($C$ = Al, Si, Sn and Sb). Using full symmetry potentials, Mohn {\it et al.} \cite{MBS95} found the magnetic ground state of Co$_2$Ti$C$ ($C$ = Al and Sn), but not a half-metallic state. Galanakis {\it et al.} \cite{GDP02} reported half-metallic behavior in various $A_2BC$ compounds, but not for the Co$_2$ compounds with Ti or Fe. The results were compatible with those found for the Mn compounds as calculated by Picozzi {\it et al.} \cite{PCF02} using the generalized gradient approximation instead of the pure LSDA. The generalized gradient approximation (GGA), as introduced by Perdew {\it et al.} \cite{PWa86,PWa92,PBE96,PBE97}, accounts for gradients of the density that are absent in the pure LSDA parameterization of the exchange-correlation functional \cite{KSh65,HLu71,BHe72,VWN80}. Using spherical potentials and the GGA, a half-metallic state could not be verified for Co$_2$FeAl \cite{MNS04,ADK05}. A half-metallic ferromagnetic ground state was also found for the complete series Co$_2$Cr$_{1-x}$Fe$_x$Al, when the full symmetry potentials were used along with the GGA in the calculations \cite{FKW05}. This clearly indicates that to find the correct ground state, electronic structure for the Heusler compounds, one may need both the full symmetry potentials and the generalized gradient approximation. With this information, the properties of the reported transition metal based Heusler compounds were calculated in the present work using both the GGA and the full symmetry potentials. However, even at that state of sophistication for some compounds the magnetic properties could not be explained properly. Therefore, the LDA+$U$ method was also used to account for on-site correlation at the transition metal sites.

For both scientific and technological reasons it is important to determine the electron spin polarization at the $\epsilon_F$ of a material, although it is difficult to measure \cite{SBO98}. The spin polarization at $\epsilon_F$ of a ferromagnet is defined by:

% EQ 1 %%%%%%%%%%%%%%%%%%%%%%%%%%%%%%%%%%%%%%%%%%%%%%%%%%%%%%%%%%%%%
\begin{equation}
     P = \frac{N_\uparrow(\epsilon_F) - N_\downarrow(\epsilon_F)}{N_\uparrow(\epsilon_F) + N_\downarrow(\epsilon_F)},
\label{eq1}
\end{equation}
%%%%%%%%%%%%%%%%%%%%%%%%%%%%%%%%%%%%%%%%%%%%%%%%%%%%%%%%%%%%%%%%%%%%

where $N_\uparrow(\epsilon_F)$ and $N_\downarrow(\epsilon_F)$ are the spin dependent density of states at the $\epsilon_F$. When the electrons are fully spin polarized, either $N_\uparrow(\epsilon_F)$ or $N_\downarrow(\epsilon_F)$ equal zero and therefore the spin polarization will be 100$\%$. Electronic band structure calculations and magnetic measurements have revealed that nearly all known Heusler compounds in the Co$_2BC$ series are ferromagnetic with a significant moment on the Co as well as on the $B$ sites.

	In this article, the classification scheme for half-metals as proposed by Coey {\it et al.} is used. The first class is Type~I, where only one type of spin polarized electrons (either up or down) contribute to the conductivity. In Type~Ia and Type~Ib the gap is in the majority and the minority densities, respectively. In Type~III half-metals, localized spin up and delocalized spin down electrons contribute to the conductivity. In Type~IIIa and Type~IIIb the series of localization and spin is interchanged.

A systematic examination of the electronic and the magnetic structure of the Heusler compounds was carried out in this work. The DOS are compared to study the effect of valence electron concentration on the magnetic properties and in particular on the minority band gap.

\subsubsection*{Slater-Pauling rule for Heusler compounds}
\label{sec:SP}

It is well known that Heusler compounds based on Co$_2$ follow the
Slater-Pauling \cite{Sla36,Pau38} rule for predicting their total spin magnetic moment.
This means that the saturation magnetization scales with the number of
valence electrons \cite{Kue00,GDP02,FKW06}. Half-metallic ferromagnets
(HMF) in Co$_2$ based Heusler compounds are supposed to exhibit a
real gap in the minority density of states and the $\epsilon_F$ is
pinned inside of the gap. From this point of view, the Slater-Pauling
rule is strictly fulfilled with

\begin{equation}
       m_{HMF}=n_V-6
\label{eq2}
\end{equation}

for the magnetic moment per atom. For ordered compounds with different kind of
atoms it is indeed more convenient to work with all atoms of the unit
cell. In the case of 4 atoms per unit cell, as in Heusler (H)
compounds, one has to subtract 24 (6 multiplied by the number of atoms)
from the accumulated number of valence electrons in the unit cell $N_V$
($s, d$ electrons for the transition metals and $s, p$ electrons for
the main group element) to find the magnetic moment per unit cell:

\begin{equation}
       m_{H}=N_V-24.
\label{eq3}
\end{equation}

with $N_V$ denoting the accumulated number of valence electrons in the
unit cell.

This rule was first noted by K\"ubler {\it et al.} \cite{Kue84} for
C1$_b$ compounds with 3 atoms per unit cell ($m_{C1_b}=N_V-18$). In
both cases the magnetic moment per unit cell becomes strictly integer
(in multiples of Bohr magnetons $\mu_B$) for HMF, what may be seen as an
advantage of the {\it valence electron rule} compared to the original
Slater-Pauling law (Eq.\ref{eq2}) even so it suggests the existence of
different laws.

%Inspecting other non Co$_2$ based Heusler compounds, one finds that 
%compounds with magnetic moments above the
%expected Slater-Pauling values are for $A$ = Fe based Heusler compounds. Those with lower values
%are either $A$ = Cu or $A$ = Ni based, with the Ni based compounds exhibiting
%higher moments compared to the Cu based compounds at the same number of
%valence electrons. Moreover, some of the Cu or Ni based compounds are
%not ferromagnetic independent of the number of valence electrons.
%Besides Mn$_2$VAl, only compounds containing both, Fe and Mn, were
%found to exhibit HMF character with magnetic moments according to the
%Slater-Pauling rule.
 
%%%%%%%%%%%%%%%%%%%%%End Section Introduction%%%%%%%%%%%%%%%%%%%%%%%%%%%%

%%%%%%% BEGIN Section CRYSTAL STRUCTURE AND COMPUTATIONAL DETAILS %%%%%%%%%%

\section{Crystal structure and calculational details}

\begin{figure}
\centering
\includegraphics[width=8cm]{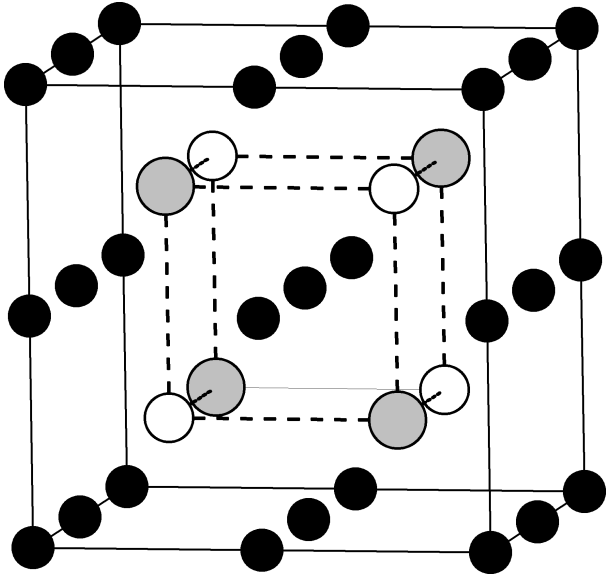}
\caption{Structure of the $A_2BC$ Heusler compounds. 
         Black $A$ atoms are at the origin and $(\frac12, \frac12, \frac12)$,
         $B$ (gray) and $C$ (white) atoms at $(\frac14, \frac14, \frac14)$ and $(\frac34, \frac34, \frac34)$,
         respectively. Note the positions are shifted by $(\frac14, \frac14, \frac14)$ with
         respect to the standard $Fm\bar{3}m$ cell to make the CsCl super structure visible.}
\label{fig1}
\end{figure}

$A_2BC$ Heusler compounds crystallize in the cubic $L2_1$ structure (space group $Fm\bar{3}m$ \cite{BRo34}). In general $A$ and $B$ atoms are transition metals and $C$ is a main group element. In some cases, $B$ is replaced by a rare earth element. The $A$ atoms are placed on 8$c$ (1/4, 1/4, 1/4) and the $B$ and $C$ atoms on 4$a$ (0, 0, 0) and 4$b$ (1/2, 1/2, 1/2) Wyckoff positions, respectively. The cubic $L2_1$ structure consists of four inter-penetrating {\it fcc} sub-lattices, two of which are equally occupied by $A$. The two $A$-site {\it fcc} sub-lattices combine to form a simple cubic sub-lattice. The $B$ and $C$ atoms occupy alternatingly the centre of the simple cubic sub-lattice resulting in a CsCl-type super structure. The crystal structure of Heusler compounds is illustrated in Fig.~\ref{fig1}.

The $\Gamma$ point of the paramagnetic structure has the symmetry $O_h$. The symmetry of the $B$ and $C$ positions is $O_h$, whereas the $A$ positions have $T_d$ symmetry. However, the wave functions at the $\Gamma$ point have to be described by $C_{4h}$ in the ferromagnetic state to account for the correct transformation of the electron spin.  
%%%%%%%%%%%%%%%%%%%%%%%%%%%%%%%%%%%%%%%%%%%
%%%%%%%%%%%%%%%%%%%%%%%%%%%%%%%%%%%%%%%%%%%
%%%%%%%%%%%%%%%%%%%%%%%%%%%%%%%%%%%%%%%%%%%
%%%%%%%%%%%%%%%%%%%%%%%%%%%%%%%%%%%%%%%%%%%
%%%%%%%%%%%%%%%%%%%%%%%%%%%%%%%%%%%%%%%%%%%

%Note: for Sir, Pearson, ABAC not respected here.............

%%%%%%%%%%%%%%%%%%%%%%%%%%%%%%%%%%%%%%%%%%%
%%%%%%%%%%%%%%%%%%%%%%%%%%%%%%%%%%%%%%%%%%%
%%%%%%%%%%%%%%%%%%%%%%%%%%%%%%%%%%%%%%%%%%%
%%%%%%%%%%%%%%%%%%%%%%%%%%%%%%%%%%%%%%%%%%%

%%%%%%%%%%%%%%%% FIGURE 1 %%%%%%%%%%%%%%%%%%%%%%%%%%%%%%%%%%%
%%%%%%%%%%%%%%%End Figure 1 %%%%%%%%%%%%%%%%%%%%%%%%%%%%%%%%

Density functional theory-based electronic structure calculations were performed using the full-potential Linear Augmented Plane Wave (FLAPW) code \textsc{Wien2k} \cite{BSM01}. The exchange-correlation functional was evaluated within the generalized gradient approximation, using the Perdew-Burke-Ernzerhof \cite{PBE96,PBE97} parameterization. The GGA specifically accounts for density gradients that are neglected in the pure LDA. The energy threshold between the core and the valence states was set to -81.6 eV. The muffin-tin-radii (R$_{MT}$) were chosen to ensure nearly touching spheres and minimizing the interstitial space. $R_{MT} \times k_{\rm max} =7$ was used for the number of plane waves and the expansion of the wave functions was set to $l = 10$ inside of the muffin tin spheres. The self-consistent calculations employed a grid of 455 $k$ points in the irreducible Brillouin zone taken from a 25$\times$25$\times$25 mesh. It turned out that this number of irreducible $k$ points is a good compromise to balance the quality of the integration and the speed of the calculation. The energy convergence criterion was set to 10$^{-5}$~Ry and the charge convergence was monitored simultaneously.
  
It was recently reported that LSDA and GGA schemes are not sufficient to describe the electronic structure correctly for Mn and Fe containing Heusler compounds based on Co$_2$ \cite{KFF06}. Here the LDA$+U$ method \cite{AAL97} was used to account for on-site correlation at the transition metal sites. The LDA+$U$ method accounts for an orbital dependence of the Coulomb and exchange interaction that is absent in the pure LDA. The effective Coulomb-exchange interaction ($U_{eff}=U-J$) was used here for the calculations. The particular LDA+$U$ self-interaction correction (SIC) scheme is used in \textsc{Wien2k} to account for double-counting corrections. The use of $U-J$ neglects, however, multipole terms in the expansion of the Coulomb interaction. It should be mentioned that the +$U$ was used on top of GGA rather than LSDA parameterization of the exchange correlation functional. No worthy differences were observed using one or the other of the parameterizations. The $U_{eff}$ values for the different elements being used for the calculations are summarized in Tab.~\ref{Tab1}.
 
%%%%%%%%%%%%%%%%%%%%%%%%%%%%%%%%%%%%%%%%%%%%%%%%%%%%%%%%%%%%%%%%%%%
\begin{table}
\caption{$U_{eff}$ values used for the different elements for LDA+$U$ calculations of Heusler compounds}
\centering
\begin{tabular}{ l c}
\hline
element  &  $U_{eff}$ (in eV) \\
Ti & 1.36 \\
V  & 1.34 \\
Cr & 1.59 \\
Mn & 1.69 \\
Fe & 1.80 \\
Co & 1.92 \\
\hline
\end{tabular}
\label{Tab1}
\end{table}
%%%%%%%%%%%%%%%%%%%%%%%%%%%%%%%%%%%%%%%%%%%%%%%%%%%%%%%%%%%%%%%%%%%%
It was found that the spin-orbit interaction has only a weak influence on the half-metallic ferromagnetism in Heusler compounds \cite{MSZ04}, therefore it was neglected in the calculations discussed here. The calculations include the mass velocity and Darwin terms to correct relativistic effects.

\section{Results and discussions}
In the following the electronic structure and magnetic properties of the Heusler compounds will be discussed in detail. Starting point are the Co$_2$ based Heusler compounds.

\subsection{Structural optimization for Co$_2$CrSi and Co$_2$ScSi}

\begin{figure}
\includegraphics[keepaspectratio, width=8cm]{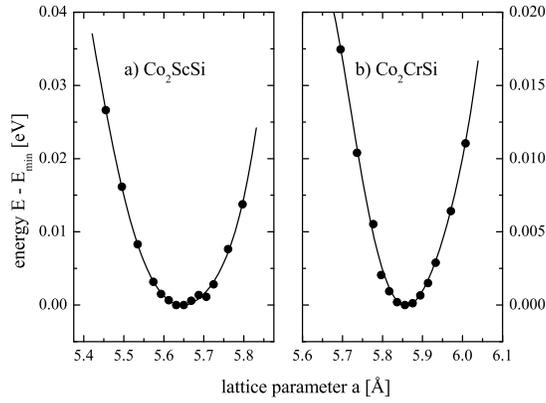}
\caption{Optimization of the lattice parameter for Co$_2$ScSi and Co$_2$CrSi compounds.
        (The solid lines result from a polynomial fit.)}
\label{fig2}
\end{figure}

It might be of interest to have first a look at those compounds which are missing (no reported yet) from the series of Co$_2BC$ Heusler compounds. The compound Co$_2$ScSi is not reported, and neither is Co$_2$CrSi. Within the FLAPW scheme, a structural optimization was performed for these two compounds. It was confirmed that the ferromagnetic configuration is lower in energy than the non spin-polarized case for both compounds. An anti-parallel spin arrangement of the Cr atoms with respect to the Co atoms in the cubic lattice showed to be energetically unfavorable with respect to the ferromagnetic arrangement in the Co$_2$CrSi compound. The results of the structural optimization are shown in Fig.~\ref{fig2}. The values of the optimized lattice parameters are given in Tab.~\ref{Tab3}. The values of the calculated total spin magnetic moment and the elemental resolved moments are also given in Tab.~\ref{Tab3}. Both compounds exhibit large values of the minority band gap. The size of the minority band gap is found to be 0.246~eV and 0.878~eV and the density of states at $\epsilon_F$ for majority states are 1.36~eV$^{-1}$ and 2.9~eV$^{-1}$ for Co$_2$ScSi and Co$_2$CrSi, respectively. The non-existence of Co$_2$CrSi and Co$_2$ScSi could be associated with the high peak in the density of states at $\epsilon_F$.

Both compounds have magnetic moments of 1 and 3~$\mu_B$ for Co$_2$ScSi and Co$_2$CrSi, respectively. This is in good agreement with the Slater-Pauling like behavior. The magnetic moment at the Co site is 1~$\mu_B$, the remainder arises from the $B$ site. The majority density of states at $\epsilon_F$ is higher in Co$_2$CrSi as compared to the Co$_2$ScSi compound, as expected. The Cr compound exhibits a larger minority gap compared to the Sc compound.

%%%%%%%%%%%%%%%%%%%%%%%%%%%%%%%%%%%%%%%%%%%%%%%%%%%%%%%%%%%%%%%%%%%%
%%%%%%%%%%%%%%%%%%%%%%%%%%%%%%%%%%%%%%%%%%%%%%%%%%%%%%%%%%%%%%%%%%%%

%%%%%%%%%%%%%%%%%%%%%%%%%%%%%%%%%%%%%%%%%%%%%%%%%%%%%%%%%%%%%%%%%%%%
\begin{figure*}
\includegraphics[keepaspectratio, width=14cm]{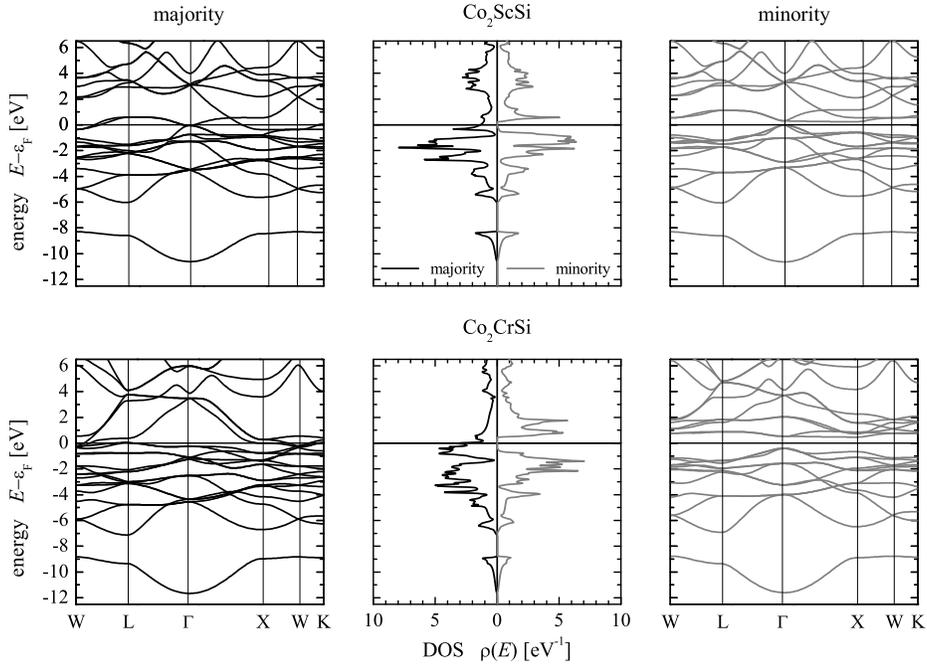}
\caption{Band structure and density of states for the hypothetical compounds Co$_2$ScSi and Co$_2$CrSi. }
\label{fig3}
\end{figure*}
%%%%%%%%%%%%%%%%%%%%%%%%%%%%%%%%%%%%%%%%%%%%%%%%%%%%%%%%%%%%%%%%%%%%

Figure~\ref{fig3} shows the calculated DOS and band structures of both hypothetical compounds. These band structures may serve as a common basis for the explanation of many Co$_2$-based Heusler compounds. They exhibit half-metallic ferromagnetic behavior with only majority bands crossing $\epsilon_F$. The band gap in minority states results in 100 \% spin polarization for these compounds.

%%%%%%%%%%%%%%%% End Section Xrystal Structure and Calculational Details%%%%%%%%%%%%%

%%%%%%%%%%%% Begin Section Results and Discussion %%%%%%%%%%%%%%%%%%%%%%%%%

\subsection{Magnetic and electronic structure}
In the half-metallic Heusler compounds discussed here, the gap stays with few exceptions in the minority spin channel, whereas $\epsilon_F$ cuts through bands in the majority spin channel. For the majority spin channel, the position of $\epsilon_F$ is in the region of the $d$ derived bands. These states are shifted to lower energies with respect to the corresponding minority spin states by the exchange splitting. Half metallic-behavior has been already predicted for a large number of Heusler compounds which are members of the series of ferromagnetic compounds. In general the $B$ atoms and in some cases also the $A$ atoms carry the magnetic moments in these compounds. It is well known that the magnetic properties of these compounds depend sensitively on the degree of atomic order and on the valence electron concentration \cite{GDP02,MNS04}. The magnetic properties depend on whether the $C$ component is a member of the $3A$ or $4A$ main group, with the latter group resulting in higher magnetic moments and Curie temperatures. Spin polarized electronic structure calculations indicate that the moments are predominantly of $3d$ origin. The density of state of the minority states is almost zero at $\epsilon_F$ whereas the majority $d$ density can have a peak or a valley close to this energy. It is filling and emptying of this peak that is assumed to produce the change in the size of the magnetic moment and the Curie temperature. In order to verify this, and to obtain further evidence for half-metallic behavior, a series of calculations for $A_2BC$ compounds was performed and is presented in the following.

%%%%%%%%%%%%%%%%%%%%%%%%%%%%%%%%%%%%%%%%%%%%%%%%%%%%%%%%%%%%%%%%%%%%%
%%%%%%%%%%%%%%%%% Hyperfine field values %%%%%%%%%%%%%%%%%%%%%%%%%%%%
%\begin{table}
%\caption{Hyperfine magnetic moment values for $A$, $B$ and $C$ elements in Heusler compounds}
%\centering
%\begin{tabular}{ l|lll|lll }
%\hline
%         & \multicolumn{3}{|c|}{GGA}                  & \multicolumn{3}{c}{LDA+$U$}\\
%\hline
%compound & H$_{hf}$(Co) & H$_{hf}$(B) & H$_{hf}$(C) & H$_{hf}$(Co) & H$_{hf}$(B) & H$_{hf}$(C) \\
%\hline
%Co$_2$TiAl & 53.182   & -48.926  & -1.283  & 11.662   & -68.121  & 0.052   \\ 
%Co$_2$TiSi & 71.096   & -68.044  & 24.421  & 21.000   & -95.000  & 32.307  \\
%Co$_2$VAl  & 65.776   &  -88.407 & -15.311 & 36.724   & -115.036 & -9.940  \\
%Co$_2$VSi  & 69.262   & -98.194  & -7.809  & 24.098   & -113.711 & -12.080 \\ 
%Co$_2$CrAl & -18.213  & -142.747 & -37.337 & -82.147  & -159.868 & -47.656 \\
%Co$_2$MnAl & -163.596 & -187.132 & -26.709 & -259.802 & -218.807 & -45.069 \\
%Co$_2$MnSi & -100.878 & -203.554 & 23.977  & -177.234 & -227.664 & 6.987   \\
%Co$_2$FeAl & -162.434 & -220.051 & -14.491 & -241.510 & -255.904 & -36.604 \\
%Co$_2$FeSi & -92.606  & -233.311 & 72.108  & -145.316 & -270.126 & 81.228  \\
%\hline 
%\end{tabular}
%\label{Tabnew}
%\end{table}
%%%%%%%%%%%%%%%%%%%%%%%%%%%%%%%%%%%%%%%%%%%%%%%%%%%%%%%%%%%%%%%%%%%%
%%%%%%%%%%%%%%%%%%%%%%%%%%%%%%%%%%%%%%%%%%%%%%%%%%%%%%%%%%%%%%%%

\subsubsection{Magnetic properties using generalized gradient approximation}

%%%%%%%%%%%%%%%%%%% BEGIN TABLE 1%%%%%%%%%%%%%%%%%%%%%%%%%
\begin{table*}
\caption{Magnetic moments of Co$_2BC$ Heusler compounds. The results of the calculations are compared to experimental values. All magnetic moments are given in $\mu_B$. $N_V$ is the number of valence electrons per formula unit (within parentheses). Total magnetic moments $m_{Calc}$ are given per unit cell. $m_A$ and $m_B$ are the site resolved magnetic moments on $A$ and $B$ sites, respectively. Experimental values of the lattice parameter $a$ and  magnetic moments $m_{exp}$ are taken mainly from references \cite{BEv81,BEJ83,EBE83,BNW00,LB19C,LB32C}, additional later references are given in the table.}
\centering
\begin{tabular}{ l|c|c|c|c|c|c|c|c|c}
\hline
         &           &           & \multicolumn{3}{|c|}{GGA}   & \multicolumn{3}{|c|}{LDA+U}   &          \\
\hline
compound ($N_V$)& $a_{exp}$ & $m_{exp}$ & $m_{calc}$ & $m_A$ & $m_B$  & $m_{calc}$ & $m_A$ & $m_B$  & Ref.\\
\hline
Co$_2$TiAl (25) & 5.847  & 0.74 & 1.00 & 0.67 & -0.18 &1.0      & 0.84  & -0.39 & \cite{KIK92}                        \\
Co$_2$TiGa (25) & 5.85   & 0.75 & 1.00 & 0.63 & -0.15 &     &   &   &  \cite{BPr75,Ooi85,SKN01}                                   \\
\hline
Co$_2$TiSi (26) & 5.743  & 1.65 & 2.00 & 1.03 & -0.02 &2.0  & 1.16  & -0.18 &                             \\
Co$_2$TiGe (26) & 5.807  & 1.59 & 1.97 & 1.05 & -0.06 & &   &   &                                         \\
Co$_2$TiSn (26) & 6.077  & 1.96 & 1.96 & 1.08 & -0.07 & &   &   & \cite{KIK92,PSS93,KKW06}                                        \\
Co$_2$VAl  (26) & 5.722  & 1.95 & 2.00 & 0.94 & 0.218 & 2.0  &1.15   & -0.09    &                         \\
Co$_2$VGa  (26) & 5.779  & 1.92 & 2.01 & 0.97 & 0.16  &     &   &   &  \cite{MGK81,CSP96}                       \\
\hline
Co$_2$VSn  (27) & 5.96   & 1.21 & 3.03 & 1.10 & 0.86  & &   &   &   \cite{FET72}                                      \\
Co$_2$VSi  (27) & 5.657  &      &  3.00  & 0.789 & 1.09  & 3.00 & 1.20  & 0.69 &  \cite{Gla62}            \\
Co$_2$CrAl (27) & 5.727  & 1.55 & 3.00 & 0.83 & 1.47  &3.01 &0.79   &1.70   &                             \\
Co$_2$CrGa (27) & 5.805  & 3.01 & 3.05 & 0.76 & 1.63  & &   &   &   \cite{UKF05}                          \\
Co$_2$CrIn (27) & 6.060  & 1.10 & 3.20 & 0.67 & 1.98  & &   &   &   \cite{WFF06}                          \\
\hline
Co$_2$MnAl (28) & 5.749  & 4.04 & 4.04 & 0.76 & 2.745 &4.28 &0.73   &3.17   &  \cite{Sol83}                           \\
Co$_2$MnGa (28) & 5.767  & 4.05 & 4.12 & 0.75 & 2.78  & &   &   &                                         \\
\hline
Co$_2$MnSi (29) & 5.645  & 4.9  & 5.00 & 1.00 & 3.00  &5.0  &0.979  &3.29   &   \cite{Ido86,Sob88,RRH02,RXJ03}                          \\
Co$_2$MnGe (29) & 5.749  & 4.93 & 5.00 & 1.02 & 3.06  & &   &   &  \cite{Ido86}                                       \\
Co$_2$MnSn (29) & 5.984  & 5.08 & 5.03 & 0.97 & 3.23  & &   &   &  \cite{Cas53,SKR72,UHl85,Ido86,Sob88,ZJM05}                                       \\
Co$_2$FeAl (29) & 5.73   & 4.96 & 4.98 & 1.23 & 2.78  &5.00 &1.22   &2.97   &  \cite{EWF04a}                           \\
Co$_2$FeGa (29) & 5.737  & 5.04 & 5.02 & 1.20 & 2.81  &     &   &   &   \cite{ZBB04,UKF05}                      \\
\hline
Co$_2$FeSi (30) & 5.64   & 6.00 & 5.59 & 1.40 & 2.87  &6.00 &1.50   &3.14   & \cite{NBH79,WFK06}                \\
Co$_2$FeGe (30) & 5.738  & 5.90 & 5.70 & 1.42 & 2.92  &     &   &   &                                     \\
\hline
\end{tabular}
\label{Tab2}
\end{table*}
%%%%%%%%%%%%%%%%%%%%%%%%%%%%%%%%%%%%%%%%%%%%%%%%%%%%%%%%%%%%%%%%%%%%

Starting with the Co$_2$-based compounds, all the information about the atom-resolved moments, total spin moments, the experimental lattice parameters, the experimental and the calculated magnetic moments are summarized in Tab.~\ref{Tab2}. For each compound, calculations were carried out using the experimental lattice parameters, and most of them exhibit at least nearly HMF type character (gap in the minority or majority states). This is clear from Figs.~\ref{fig6} and \ref{fig7} where the spin projected density of states of nine selected compounds are shown (note: details will be discussed later). Those will be used as representatives of the compounds as listed in Tab.~\ref{Tab2}. These nine compounds are Co$_2BC$ Heusler compounds with $B$ = Ti, V, Cr, Mn, Fe and $C$ = Al, Ga and Si. By making such a choice, one covers both a range of electronically different kinds of $B$ and $C$ atoms with Co atoms at the $A$ site. In most of the cases the total spin magnetic moment is exactly integer as expected for a half-metallic system. The experimental magnetic moments are also given in Tab.~\ref{Tab2} for comparison with calculations. In most cases, the calculated magnetic moments are in good agreement with the experimental results. $m_{cal}$ is the calculated total spin magnetic moment of the compound, which is the combination of the moments at $A$ site (2 times), the $B$ sites, the $C$ sites and the moment of the interstitial between the sites. That is the reason why $m_A$ and $m_B$ alone are not summing up to result in $m_{cal}$. The missing or excess of the total moment is found at the $C$ sites and in the interstitial between the sites. Exceptions from experimental values appear for compounds carrying a high magnetic moment like Co$_2$FeSi, Co$_2$FeGe etc. Inspection of the magnetic moment revealed that the Co and $B$ atoms possess high spin magnetic moments in this series of compounds. In case of small magnetic moment compounds, the Co atoms contribute most to the moment compared to the compounds with large magnetic moments. While going from the low to the high magnetic moment side, the $B$ atoms contribute an increasing moment in these compounds.  

In all reported compounds, the $C$ atoms carry a negligible magnetic moment, that does not contribute much to the overall moment. In most of the compounds it is aligned anti-parallel to the $A$ and $B$ moments. It emerges from the overlap of the electron wave functions.

There are some cases where simple LSDA-GGA does not give the correct magnetic moments, and $\epsilon_F$ may not fall into the minority gap. The magnetic moment of most of the compounds in this series exhibits a linear behavior with the number of valence electrons and follows the Slater-Pauling rule \cite{Kue00,GDP02,FKW06}. Note that there are some deviations from this trend: for example, Co$_2$FeAl, Co$_2$VGa, Co$_2$TiGa and Co$_2$ScSi.

%%%%%%%%%%%%%%%%%%%%%%%%%%%%%%%%%%%%%%%%%%%%%%%%%%%%%%%%%%%%%%%%%%%%

%%%%%%%%%%%%%%%%%%%%%%%%%%%%%%%%%%%%%%%%%%%%%%%%%%%%%%%%%%%%%%%%%%%%
%%%%%%%%%%%%%%%%%%%%%%%%%%%%%%%%%%%%%%%%%%%%%%%%%%%%%%
%%%%%%%%%%%%%%%%%%%%%%%%%%%%%%%%%%%%%%%%%%%%%%%%%%%%%%
%%%%%%%%%%%%%%%%%%%%%%%%%%%%%%%%%%%%%%%%%%%%%%%%%%%%%%
%Note: For Sir- Note differences to C1b compounds.

%%%%%%%%%%%%%%%%%%%%%%%%%%%%%%%%%%%%%%%%%%%%%%%%%%%%%%
%%%%%%%%%%%%%%%%%%%%%%%%%%%%%%%%%%%%%%%%%%%%%%%%%%%%%%
%%%%%%%%%%%%%%%%%%%%%%%%%%%%%%%%%%%%%%%%%%%%%%%%%%%%%%
%%%%%%%%%%%%%%%%%%%%%%%%%%%%%%%%%%%%%%%%%%%%%%%%%%%%%%

\subsubsection{Magnetic behavior with valence electron concentration}
\label{sec:MVEC}

%%%%%%%%%%%%%%%%%%%%%%%%%%%%%%%%%%%%%%%%%%%%%%%%%%%%%%%%%%%%%%%%%%%%

\begin{table}
\caption{Optimized lattice parameters (in \AA), total magnetic moments, and atom resolved magnetic moments (in $\mu_B$).
 $\Delta(a)$ is the change of the lattice parameter (in \%) with respect to the experimental value}
\centering \begin{tabular}{ l| c | c | c | c | c |}
\hline
Compound  & a$_{opt}$ & $\Delta(a)$ & m$_{tot}$ & m$_A$ & m$_B$ \\
\hline
Co$_2$ScAl &5.960 &      & 0 & 0.00 &  0.00 \\
Co$_2$TiAl &5.828 & -0.3 & 1 & 0.62 & -0.13 \\
Co$_2$VAl  &5.754 & +0.6 & 2 & 0.94 &  0.23 \\
Co$_2$CrAl &5.708 & -0.3 & 3 & 0.80 &  1.52 \\
Co$_2$MnAl &5.695 & -0.9 & 4 & 0.77 &  2.67 \\
Co$_2$FeAl &5.692 & -0.7 & 5 & 1.22 &  2.79 \\
\hline
\hline
Co$_2$ScSi &5.865 &      & 1    & 0.60 & -0.10 \\
Co$_2$TiSi &5.760 & +0.3 & 2    & 1.03 & -0.02 \\
Co$_2$VSi  &5.688 & +0.5 & 3    & 1.10 &  0.80 \\
Co$_2$CrSi &5.647 &      & 4    & 1.00 &  2.03 \\
Co$_2$MnSi &5.643 & 0    & 5    & 1.06 &  2.99 \\
Co$_2$FeSi &5.634 & 0    & 5.55 & 1.39 &  2.85 \\
\hline
\end{tabular}
\label{Tab3}
\end{table}
%%%%%%%%%%%%%%%%%%%%%%%%%%%%%%%%%%%%%%%%%%%%%%%%%%%%%%%%%%%%%%%%%%%%

%%%%%%%%%%%%%%%%%%%%%%%%%%%%%%%%%%%%%%%%%%%%%%%%%%%%%%%%%%%%%%%%%%%%
\begin{figure}
\includegraphics[keepaspectratio, width=8cm]{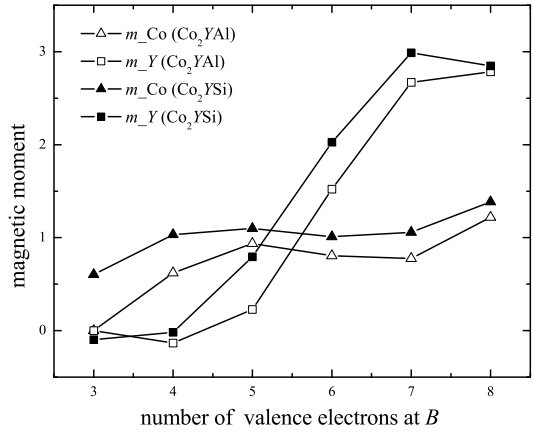}
\caption{Evaluation of the magnetic moments of Co and $B$ atoms in the series Co$_2BC$ ($C$ = Al, Si).}
\label{fig4}
\end{figure}
%%%%%%%%%%%%%%%%%%%%%%%%%%%%%%%%%%%%%%%%%%%%%%%%%%%%%%%%%%%%%%%%%%%

In this subsection, the magnetic behavior of different compounds with different valence electron concentrations is discussed. The structure of the two series of compounds Co$_2B$Al and Co$_2B$Si ($B$ = Sc, Ti, V, Cr, Mn, Fe), were first geometrically optimized. The energy minimum which defines the optimal $a$ was taken as input for further calculations. The optimized lattice parameters were found to be in average within 1\% of the experimentally obtained lattice parameters. In the two cases, Co$_2$MnSi and Co$_2$FeSi, the optimized lattice parameter was found to match the experimental one. The optimized lattice parameters are given in Tab.~\ref{Tab3} together with the total and element specific magnetic moments for Co and $B$ sites.

In the Co$_2B$Al series, the magnetic moment at the Co site first increases and then stays at about 1 $\mu_B$, while the magnetic moment at the $B$ site starts from small negative values and then increases linearly with the valence electron concentration (see in Figure~\ref{fig4}). In all compounds, the magnetic moment at the Al and Si sites is very small and anti-parallel to Co. It seems that the moment of Co is fixed at about 1$\mu_B$ and compels the $B$ moment to result in the overall magnetic moment according to the valence electron concentration.

The magnetic moment at the Co site stays at about 1 $\mu_B$ from Ti to Mn and diverges from this value only for Sc and Fe containing compounds. The magnetic moment at the $B$ site increases linearly from Ti to Mn compounds such that the total magnetic moment follows the Slater-Pauling rule. With an increasing number of valence electrons, the magnetic moment increases linearly up to Mn and after that it deviates from this trend. At least for high magnetic moment compounds one should consider on-site electron-electron correlation. In the Co$_2B$Si series, the trends are the same as in the previously discussed Al series. The only difference is the detail of the magnetic moments at Co and $B$ atoms. The increase of the magnetic moments is attributed to the rearrangement of the electrons (discussed briefly in the next section). In summary the Co atoms contribute 1~$\mu_B$ and {\it drive} the properties of the $B$ atoms to have a magnetic moment according to the number of valence electrons of the compounds. The properties of the Co$_2BC$ compounds are dominated by the nature of the Co atoms which force the {\it extent of localization} of the electrons and the resulting magnetic moment at the $B$ site.

\subsection{Electronic structure and density of states}
\label{sec:BS}
As seen in the previous subsection, the Co$_2BC$ compounds can be distinguished in two classes, one with small magnetic moments (less than 3~$\mu_B$) and the remainder with high magnetic moments. The class of compounds with small magnetic moments will be called ``low-$m$" and the one with high magnetic moments ``high-$m$". In the next subsection, the evaluation of the band gap in the minority states is discussed.
  
\subsubsection{Band gap}

%%%%%%%%%%%%%%%%%%%  BEGIN TABLE 2 %%%%%%%%%%%%%%%%%%%%%%%%%%%%%%%%%
\begin{table*}
\caption{Co$_2BC$ compounds with minority band gaps including $\epsilon_F$ (Type~I half-metal). $E_{min}$ and $E_{max}$ are the values of the minimum energy of the conduction band and the maximum energy of the valence band. The band gap $\Delta E$ is the difference between these external energies. All energies are given in eV. $N_{\uparrow}(\epsilon_F)$ is the density of states at $\epsilon_F$ for majority electrons.The last four compounds are Type~III half metallic ferromagnets.}
\centering
\begin{tabular}{l | c | c | c | c |c | c | c}
\hline
           & \multicolumn{4}{|c|}{GGA}           & \multicolumn{3}{|c}{LDA+U}                        \\
\hline
Compound   & $E_{max} (\Gamma)$ & $E_{min}$ ($A$) & $\Delta E$ & $N_{\uparrow}(\epsilon_F)$ & $E_{max}$ & $E_{min}$ & $\Delta E$  \\
\hline
Co$_2$TiAl & -0.24  & 0.215   & 0.456 & 1.37  & -0.206 & 0.914 & 1.120   \\

Co$_2$TiSi & -0.606 & 0.179   & 0.785 & 0.986 & -0.583 & 0.632 & 1.215   \\
Co$_2$TiGe & -0.401 & 0.194   & 0.595 & 0.93  &        &       &         \\
Co$_2$TiSn & -0.222 & 0.282   & 0.504 & 1.05  &        &       &         \\
Co$_2$VGa  & -0.142 & 0.047   & 0.189 & 1.52  &        &       &         \\
Co$_2$VSn  & -0.397 & 0.151   & 0.548 & 5.34  &        &       &         \\
Co$_2$CrAl & -0.118 & 0.63    & 0.748 & 4.84  & 0.071  & 1.39  & 1.319   \\
Co$_2$MnSi & -0.292 & 0.506   & 0.798 & 1.27  & 0.007  & 1.307 & 1.300   \\
Co$_2$MnGe & -0.048 & 0.533   & 0.581 & 1.29  &        &       &         \\
\hline
Co$_2$VSi  & -0.886 & -0.067  & 0.072 & 3.50  & -0.905 & 0.138 & 1.04     \\
Co$_2$FeAl & -0.138 & -0.027  & 0.111 & 0.87  & 0.022  & 0.811 & 0.789   \\
\hline
Co$_2$TiGa & 0.06   & 0.216   & 0.210 & 1.44  &        &       &         \\
Co$_2$FeGa & 0.086  & 0.107   & 0.021 & 0.88  &        &       &         \\
\hline
\end{tabular}
\label{Tab4}
\end{table*}
%%%%%%%%%%%%%%%%%%%%%%%%%%%%%%%%%%%%%%%%%%%%%%%%%%%%%%%%%%%%%%%%%%%%

\begin{table*}
\caption{Band gaps in the minority states of Co$_2BC$ compounds with $\epsilon_F$ outside of the gap. These are Type~III half-metallic ferromagnets. In this class of compounds the minority gap does not include $\epsilon_F$. The compounds are grouped for the cases where $\epsilon_F$ from GGA calculations is below or above the gap. (For quantities see Table~\ref{Tab2})}
 \centering \begin{tabular}{ l| c | c | c | c | c | c | c | c}
\hline
         & \multicolumn{5}{c}{GGA}            & \multicolumn{3}{|c}{LDA+U}          \\
\hline
Compound & $E_{max}$ & $E_{min}$ & $\Delta E$ &  $N_{\uparrow}$($\epsilon_F$) & $N_{\downarrow}$($\epsilon_F$) & $E_{max}$ & $E_{min}$ & $\Delta E$ \\
\hline
Co$_2$CrGa & 0.203  & 0.628  & 0.425  & 2.8  & 0.2  &        &       &       \\
Co$_2$CrIn & 0.450  & 0.612  & 0.162  & 2.35 & 0.5  &        &       &       \\
Co$_2$MnAl & 0.256  & 0.882  & 0.626  & 1.05 & 0.2  & 0.424  & 1.505 & 1.081 \\
Co$_2$MnGa & 0.362  & 0.731  & 0.369  & 1.7  & 0.34 &        &       &       \\
Co$_2$MnSn & 0.183  & 0.594  & 0.411  & 1.22 & 0.16 &        &       &       \\
\hline
Co$_2$VAl  & -0.357 & -0.119 & 0.238  & 1.6  & 0.03 & -0.414 & 0.072 & 0.486 \\
Co$_2$FeSi & -0.735 & -0.589 & 0.146  & 2.7  & 0.71 & -0.81  & -0.028& 0.782 \\
Co$_2$FeGe & -0.517 & -0.43  & 0.087  & 2.3  & 0.74 &        &       &       \\
\hline
\end{tabular}
\label{Tab5}
\end{table*}
%%%%%%%%%%%%%%%%%%%%%%%%%%%%%%%%%%%%%%%%%%%%%%%%%%%%%%%%%%%%%%%%%%%%

Table~\ref{Tab4} summarizes the results calculated for the minority band gap of the Co$_2BC$ compounds with a clear gap at $\epsilon_F$. All of the listed compounds show a gap along $\Gamma-X$ that is in the $\Delta$-direction. For easier comparison we looked along $\Gamma-X$ that is the $\Delta$-direction of the paramagnetic state. The $\Delta$-direction is perpendicular to the Co$_2$ (100)-planes. As was shown earlier \cite{FKW05}, just the $\Delta$-direction plays the important role for understanding of the HMF character and magnetic properties of Heusler compounds. This fact was also pointed out by \"O\^g\"ut and Rabe \cite{ORa95}. The listed compounds (except the four shown at the bottom of the table) exhibit a gap at $\epsilon_F$ and behave like HMF within GGA. The band gap values for some of the compounds using LDA+$U$ method are also listed in the Tab.~\ref{Tab4}. It is clear that electron-electron correlation may create or destroy the gap or shift $\epsilon_F$ outside the gap.
  
As one approaches towards high magnetic moment compounds, there is a reduction in the density of states of the majority spin states at $\epsilon_F$. Only few compounds have a high majority DOS at the $\epsilon_F$ whereas, the remainder have small values. For some high magnetic moment compounds it is completely different. Instead of having a high majority DOS they have only a small DOS. Already the GGA calculations reveal the small DOS at $\epsilon_F$. The important conclusion one can draw at this point is that the LSDA-GGA does not estimate the gap and magnetic moment of these compounds correctly and results in a Type~III half-metal.

The minority band gaps of Co$_2BC$ compounds, which are Type~III in GGA are shown in Tab.~\ref{Tab5}. Their DOS values at $\epsilon_F$ for majority and minority states are also listed in the table.

\begin{figure*}
\begin{tabular}{ll}
(a)                                                    & (b)                                                    \\
\includegraphics[keepaspectratio, width=8cm]{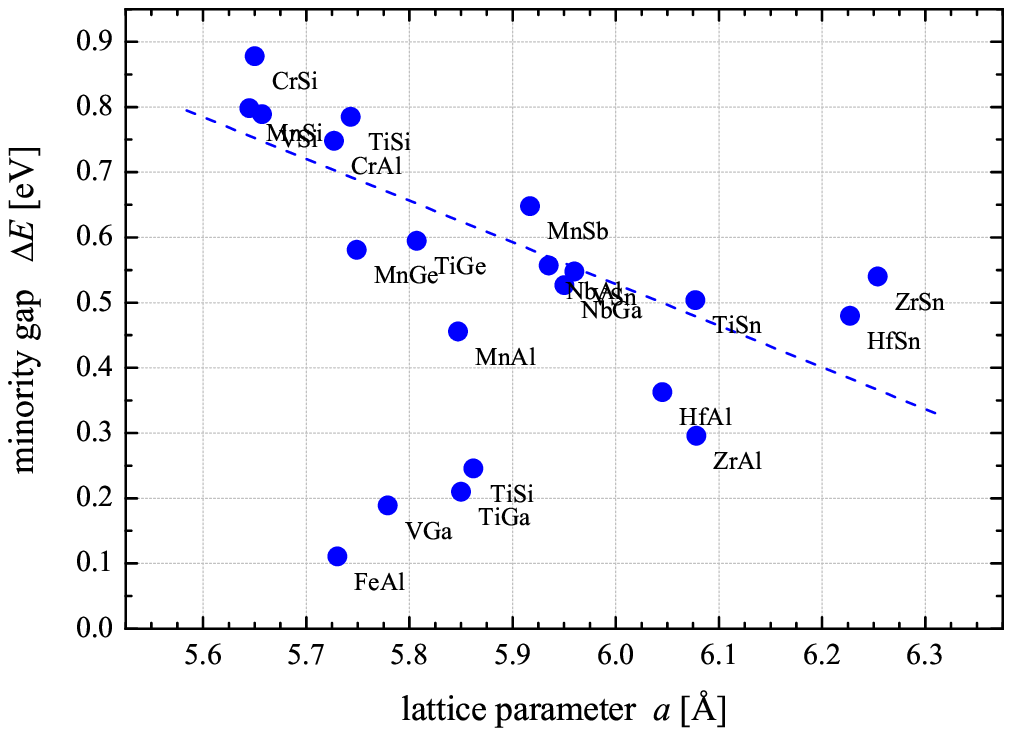} & \includegraphics[keepaspectratio, width=8cm]{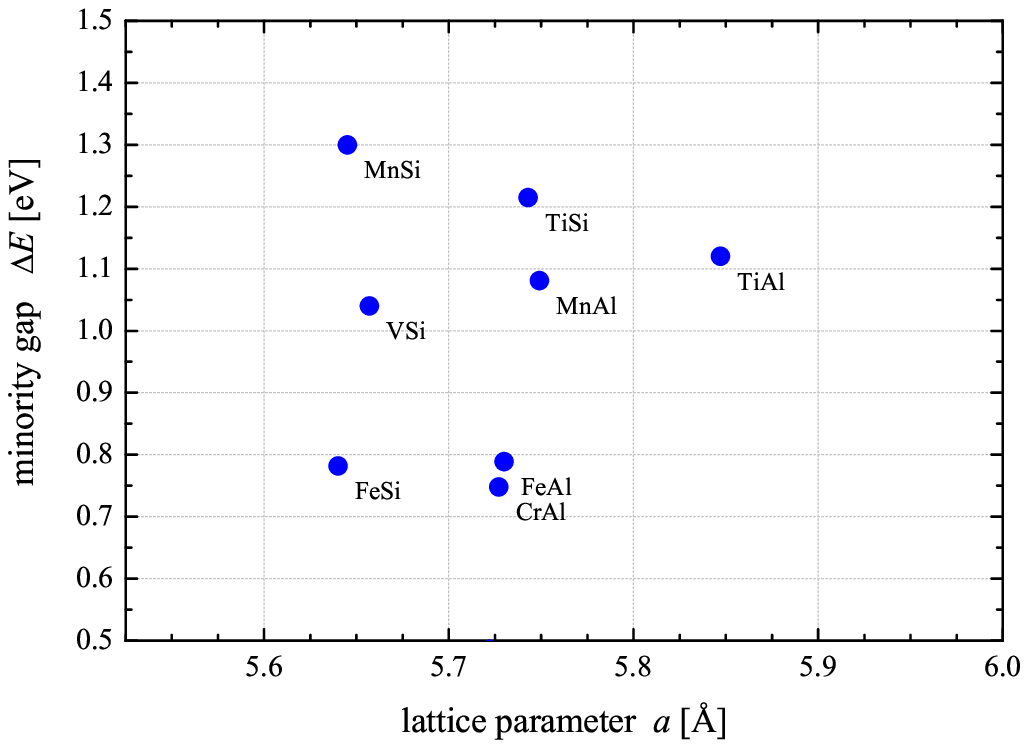}\\
\end{tabular}
\caption{Minority band gap in Co$_2$ based half-metallic ferromagnets within (a) GGA and (b) LDA+$U$. \\
         Shown is the size of the gap ($\Delta E$) versus lattice parameter ($a$).
         The line is drawn to guide the eye.}
\label{fig5}
\end{figure*}

The lattice parameter dependence as a function of the size of the minority gap is displayed in Fig.~\ref{fig5} (a) for GGA and (b) for LDA+$U$. There is a tendency that larger lattice parameter of the compounds lead to smaller band gaps. This trend cannot be used quantitatively, but is useful as a starting guess to search for good candidates for HMF materials.

%%%%%%%%%%%%%%%%%%%%%%%%%%%%%%%%%%%%%%%%%%%%%%%%%%%%%%%%%%%%%%%%%%%%
%%%%%%%%%%%%%%%%%%%%%%%%%%%%%%%%%%%%%%%%%%%%%%%%%%%%%%%%%%%%%%%%%%%

\subsubsection{Low $m$ compounds}

\begin{figure}
\includegraphics[keepaspectratio, width=8cm]{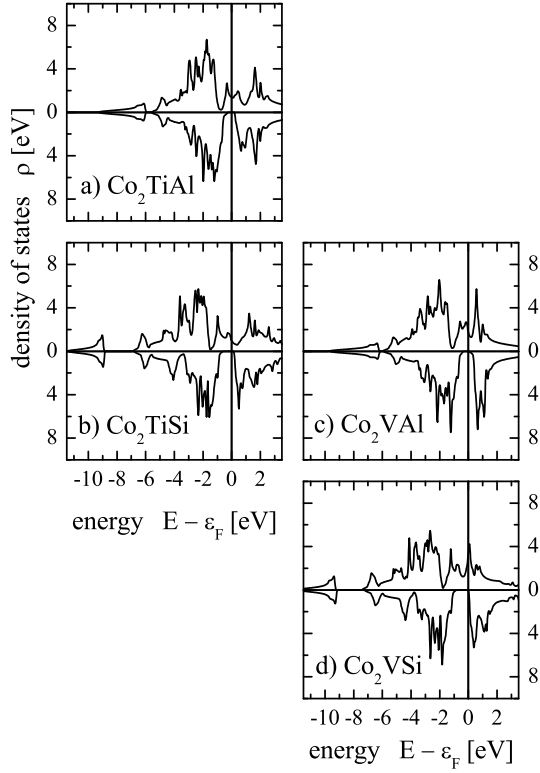}
\caption{Density of states of selected low $m$ Co$_2BC$ compounds. 
         Shown are the spin resolved DOS for compounds with $B$ = Ti, V and $C$ = Al, Si.
         (Note: upper and lower parts of the panels show for each compound
         the majority and minority DOS, respectively.)}
\label{fig6}
\end{figure}

The DOS for the low $m$ compounds Co$_2$TiAl, Co$_2$TiSi, Co$_2$VAl and Co$_2$VSi are displayed in the different panels of Fig.~\ref{fig6}. The upper part of each panel displays the majority spin states and the lower one the minority spin states.

%%%%%%%%%%%%%%%%%%%%%%%%%%%%%%%%%%%%%%%%%%%%%%%%%%%%%%%%%%%%%%%%%%%%
%%%%%%%%%%%%%%%%%%%%%%%%%%%%%%%%%%%%%%%%%%%%%%%%%%%%%%%%%%%%%%%%%%%%

The large splitting of the minority states in the Co$_2B$Si compounds is remarkable on going from lower to higher valence electron concentration. This behavior is expected for such compounds due to the higher moments at Co and $B$ sites compared to Co$_2$TiAl and Co$_2$TiSi. The majority and minority $d$ states spread up to -6~eV below $\epsilon_F$ in Co$_2$TiAl. For Co$_2$TiSi and Co$_2$VSi, the band width is almost the same and the density of the $d$ states spreads up to -7~eV below $\epsilon_F$ due to the larger crystal field splitting as well as larger exchange splitting in both compounds. Indeed, in all four compounds the states at and close to $\epsilon_F$ are strongly spin polarized and all four systems exhibit a gap in the minority states. The spin polarization is found to be 100~\% which characterizes the systems to be half-metallic.

%The DOS of the minority states is zero at $\epsilon_F$ whereas the
%majority of $d$ bands have a small peak near to this energy. It is the
%filling or emptying of majority states that is responsible for the change 
%of the magnetic moment.

% and the Curie temperature as well as
%the C component changes. Since exchange splitting of the Mn $3d$
%band is extremely large which might be the explanation for the high
%magnetic moment of Co in Co$_2$MnC compounds.

\subsubsection{High $m$ compounds}

\begin{figure}
\includegraphics[keepaspectratio, width=8cm]{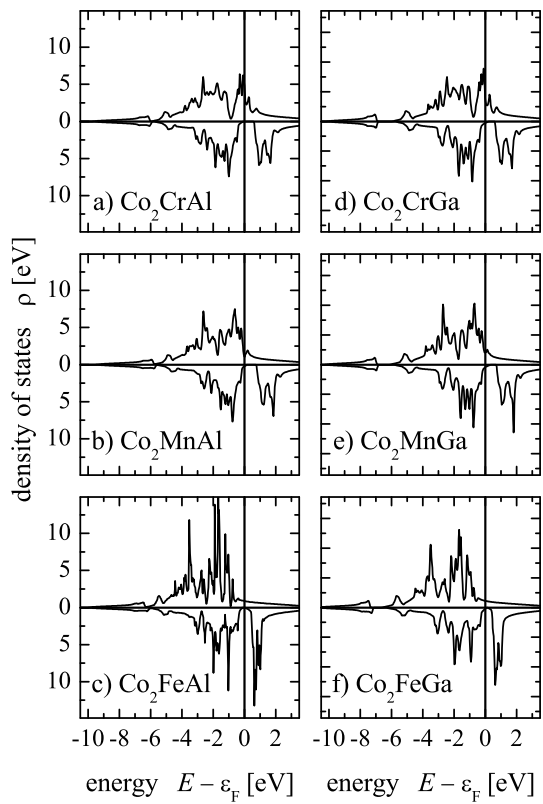}
\caption{Density of states for low to high $m$ Co$_2BC$ compounds. 
         Shown are the spin resolved DOS for compounds with $B$ = Cr, Mn, Fe and $C$ = Al, Ga.
         (Note: upper and lower parts of the panels show for each compound
         the majority and minority DOS, respectively.)}
\label{fig7}
\end{figure}

In the next step, the six compounds Co$_2BC$ ( $B$ = Cr, Mn, Fe and $C$ = Al, Ga) with high $m$ were chosen. The DOS for these compounds are displayed in the various panels of Fig.~\ref{fig7}. The situation is somewhat different from the low $m$ (as explained in the previous subsection). All listed compounds exhibit almost the same band width. However, there are distinct differences in the shapes of the states. There are some minority states appearing near $\epsilon_F$, which cause the decrease of the total spin polarization for these compounds. Again there is a larger splitting in case of Co$_2$FeGa, due to Co and Fe atoms which carry the high magnetic moment. The FLAPW spin polarized results predict an {\it approximate} half-metal behavior. The most interesting conclusion that can be drawn from the DOS of these Co$_2BC$ based Heusler compounds is that the states at and near $\epsilon_F$ are strongly spin polarized and the systems indeed exhibit half-metallic ferromagnetism. There are changes in the DOS while going from the low $m$ compounds to the high $m$. A decrease in the contribution from majority states just at $\epsilon_F$ is observed. The low density in the majority states close to $\epsilon_F$ is nevertheless arising from $d$-electrons in strongly dispersing bands. 

It is very often observed that the Ga containing Heusler compounds based on Co$_2$ have a smaller gap at $\epsilon_F$ as compared to others. These compounds may have some minority states appearing close to $\epsilon_F$.

%%%%%%%%%%%%%%%%%%%%%%%%%%%%%%%%%%%%%%%%%%%%%%%%%%%%%%%%%%%%%%%%%%%%

%%%%%%%%%%%%%%%%%%%%%%%%%%%%%%%%%%%%%%%%%%%%%%%%%%%%%%%%

\subsubsection{Distribution of the electrons}

On careful examination of the magnetic moments on the Co sites in both series (Co$_2BC$ with $C$ = Al and Si), one finds that the Co site is carrying an average of 0.8 $\mu_B$ (Al series) and 1.0 $\mu_B$ (Si series). At this point it is not clear why the magnetic moment of Co is changing with exchange of Al by Si. Thus the question arises how $C$ affects the overall magnetic moment without directly contributing to it? To check this, it is necessary to look at the electron distribution in the system. A systematic study has been carefully carried out for these two series of compounds.

\begin{figure*}
\includegraphics[keepaspectratio, width=14cm]{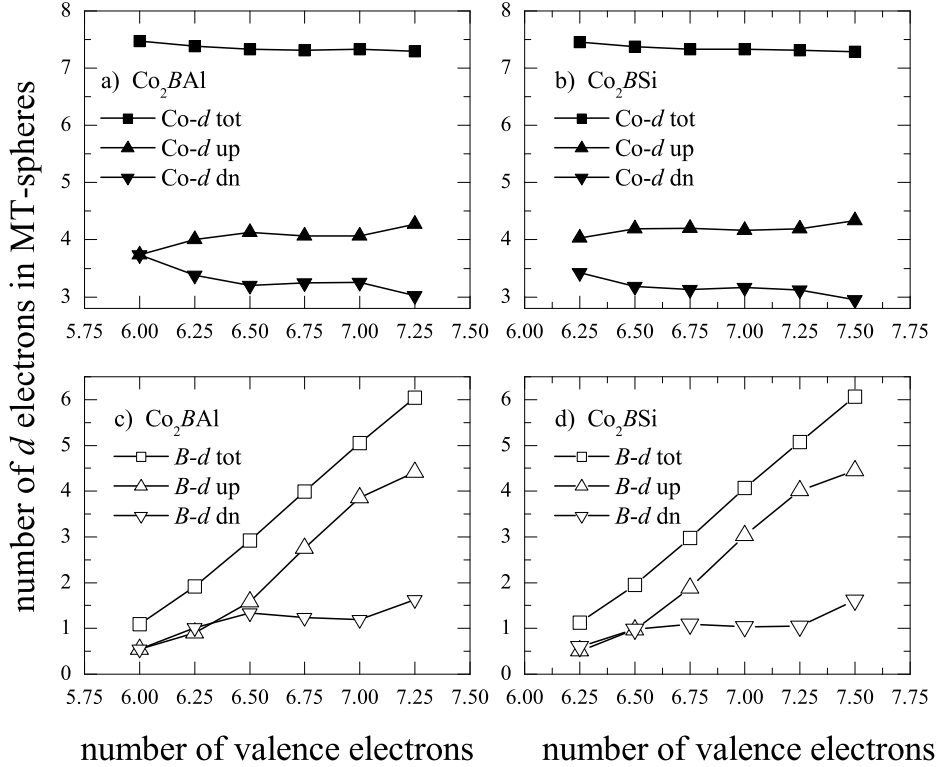}
\caption{Total number of valence $d$ electrons present in spin up and spin down channel for Co and $B$ sites in Co$_2B$Al and Co$_2B$Si; ($B$ = Sc, Ti, V, Cr, Mn, and Fe) with the average number of valence electrons.}
\label{fig9}
\end{figure*}

The total magnetic moment in the series is defined by the number of unoccupied $d$ electrons in the majority and minority states. The total number of $d$ electrons at the Co site is about 7.5 including majority and minority electrons (see Fig.~\ref{fig9}). The Co $d$ states are split up into two states for majority and minority spin. The majority and minority spin $d$ states are almost constant and contain 4 and 3.5 electrons, respectively. Their values are nearly constant and do not change much if going from Ti to Mn. This is one reason why the magnetic moment of Co is fixed at about 1 $\mu_B$.

The distribution of the $d$ electrons in majority and minority states for the $B$ element is shown in Figs.~\ref{fig9}(c, d). It is clear that filling of both majority and minority $d$ electrons increases in parallel up to V and thereafter the electrons start to fill more majority states as compared to minority states. Al and Si contribute very less to the $d$ states and in-fact they do not contribute directly to the overall magnetic moment.

%%%%%%%%%%%%%%%%%%%%%%%%%%%%%%%%%%%%%%%%%%%%%%%%%%%%%%%%%%%%%%%%%%%%
%%%%%%%%%%%%%%%%%%%%%%%%%%%%%%%%%%%%%%%%%%%%%%%%%%%%%%%%%%%%%%%%%%%%

%%%%%%%%%%%%%%%%%%%%%%%%%%%%%%%%%%%%%%%%%%%%%%%%%%%%%%%%%%%%%%%%%%%%
\begin{figure}
\includegraphics[keepaspectratio, width=8cm]{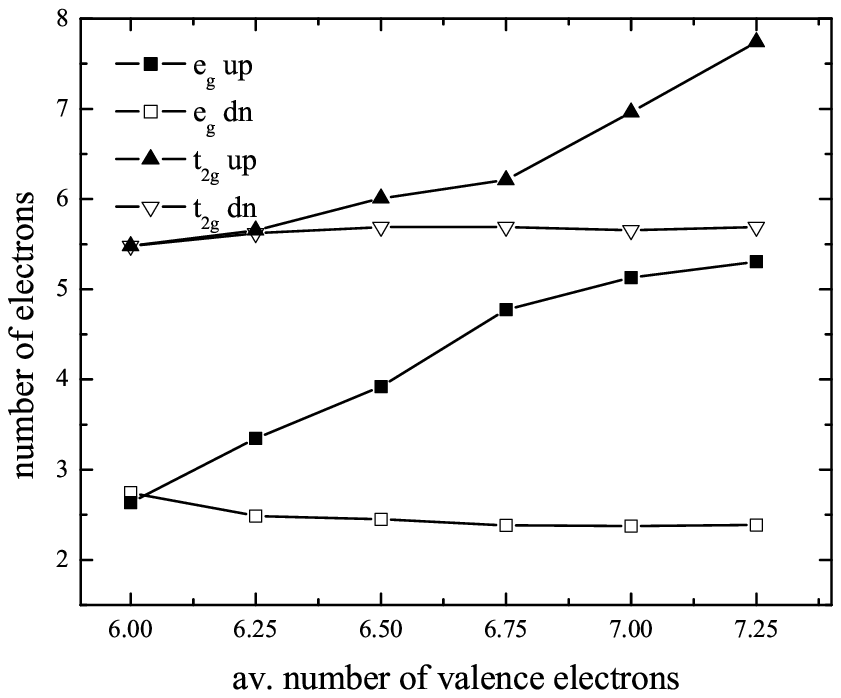}
\caption{Distribution of the electrons in symmetry distinguished states (e$_g$ and t$_{2g}$) in Co$_2B$Al compounds.}
\label{fig10}
\end{figure}
%%%%%%%%%%%%%%%%%%%%%%%%%%%%%%%%%%%%%%%%%%%%%%%%%%%%%%%%%%%%%%%%%%%%

%%%%%%%%%%%%%%%%%%%%%%%%%%%%%%%%%%%%%%%%%%%%%%%%%%%%%%%%%%%%%%%%%%%%
\begin{figure}
\includegraphics[keepaspectratio, width=8cm]{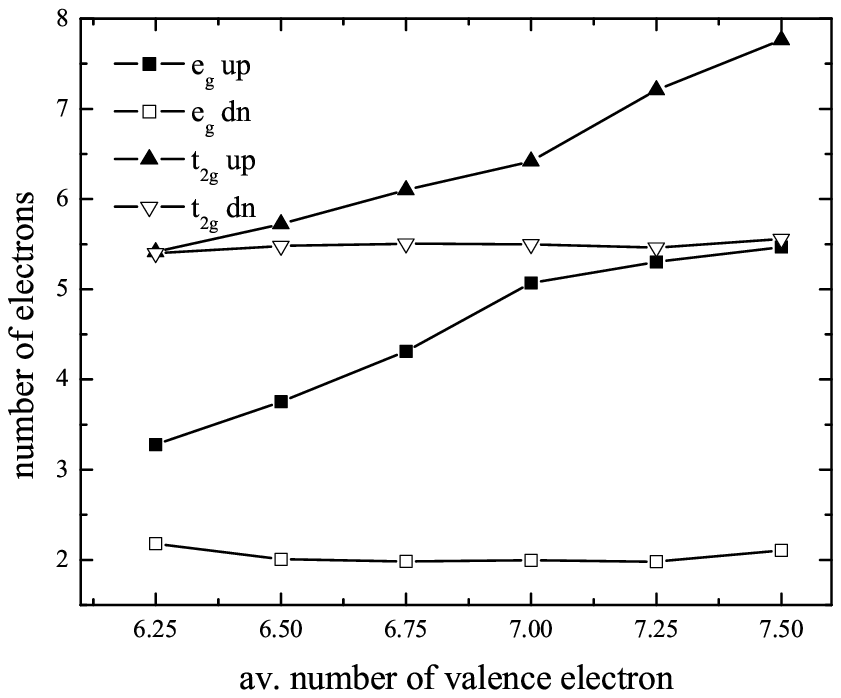}
\caption{Distribution of the $d$ electrons in symmetry distinguished states (e$_g$ and t$_{2g}$) in Co$_2B$Si compounds.}
\label{fig11}
\end{figure}
%%%%%%%%%%%%%%%%%%%%%%%%%%%%%%%%%%%%%%%%%%%%%%%%%%%%%%%%%%%%%%%%%%%%

\subsubsection{Effect of main group elements}

To check how the different main group elements ($C$) affect the magnetic moments and the number of electrons in $d$ states, the Co$_2BC$ compounds with $B$ = Sc, Ti, V, Cr, Mn, Fe; $C$ = Al and Si were studied. It was found that the trends in the Co$_2B$Si compounds are the same as for Co$_2B$Al.

At this point the question arises why the total magnetic moment increases by 1 $\mu_B$ when Al is replaced by Si. Taking simple examples like Co$_2$MnAl and Co$_2$MnSi, the first compound has a total magnetic moment of 4 $\mu_B$ and the second 5 $\mu_B$. Al and Si do not contribute directly to the total magnetic moment. In order to understand the increase of the moment, the distribution of electrons in states with different symmetry is studied in the series of Co$_2B$Al and Co$_2B$Si.

As determined earlier by K\"ubler \cite{Kue84}, to maximize the number of unpaired electrons, the minority states should be filled by 8 $d$ electrons and the rest of the electrons fills the majority states.

In the Co$_2B$Al series, overall (average) 7.85 $d$ electrons are in the minority states. There should be 8 $d$ electrons to fill the $d$-states completely and to have a gap. The missing part of the electrons is found in the interstitial. That is, they are completely delocalized and cannot be attributed to a particular atom. When inspecting the Co$_2B$Si series, the number of total minority $d$ electrons is in average 7.5. the remaining are found in the interstitial and overlap with the Si $s$ and $p$ states. To proof this situation in detail, LMTO-ASA \cite{JAn00} calculations were performed to obtain the crystal orbital Hamiltonian population (COHP) of Co-Al and Co-Si. It was found that the bonding interaction between Co and Si is much stronger than between Co and Al. This is expected from the higher electronegativity of Si compared to Al. Due to the stronger bonding interaction between Co and Si in Co$_2B$Si, some more electrons are in between the atoms and missing from the expected total count of 8. At the same time, the electrons in the majority states are redistributed to include more electrons and as result a higher magnetic moment.

Replacement of Al by Si or other members of the $3A$ and $4A$ group, plays an important role for the distribution of electrons in the various symmetry distinguished states ($t_{2g}$ and $e_g$) at Co as well as at $B$ sites, as shown in Figs.\ref{fig10} and \ref{fig11}. In-fact the overall number of $d$ electrons remains the same. Addition of an extra electron by replacing Al by Si affects mostly all symmetry distinguished states except the $e_g$ states at the $B$ site. The symmetry resolved $e_g$ states of $B$ are not affected by other states because there is no possible direct overlap to other states, whereas the $t_{2g}$ states form the bonds with the atoms at the $A$-sites. The $e_g$ states are mainly responsible for the localized magnetic moment at the $B$ sites.

\subsection{Electronic structure within LDA+$U$}

\subsubsection{High $m$ compounds and LDA$+U$}

\begin{figure*}
\centering
\includegraphics[width=14cm]{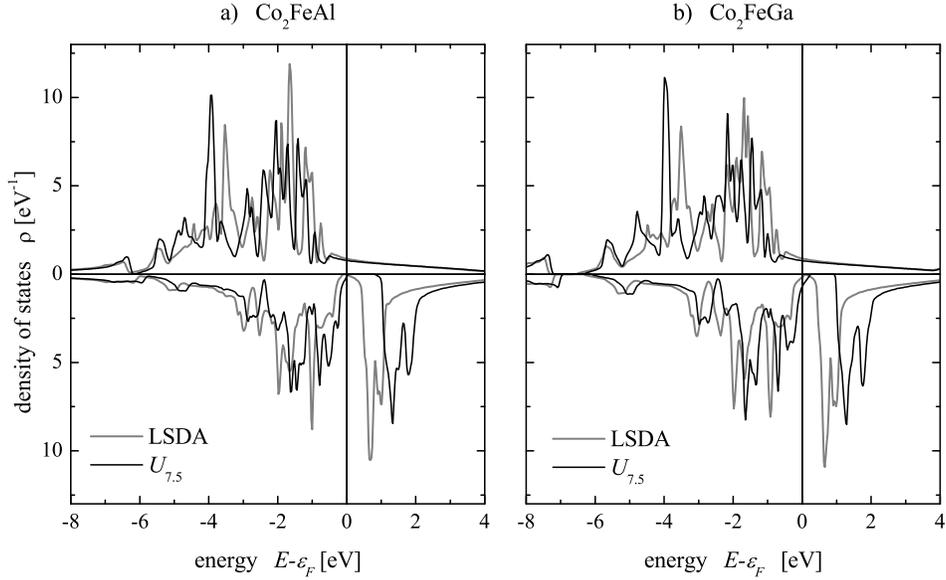}
\caption{Spin-resolved density of states for Co$_2$FeAl and Co$_2$FeGa. 
         Black lines indicate the DOS using the LDA$+U$ with $U_{7.5}$ for Co$_2$FeAl (a)
         and $U_{7.5}$ for Co$_2$FeGa (b). Gray lines indicate the results from the LSDA.
         The upper and lower parts of the plots display the majority and minority states,
         respectively. (See the text for the particular values of $U$ at the different sites.)}
\label{fig8}
\end{figure*}

In this subsection, the influence of correlation on the electronic structure of various compounds is discussed and the results are displayed in Tabs.~\ref{Tab2}, \ref{Tab4} and \ref{Tab5}. Out of these, only the band structures of the two compounds Co$_2$FeAl and Co$_2$FeGa will be discussed. These two compounds are selected as an example. The spin-resolved density of states for Co$_2$FeAl and Co$_2$FeGa that are calculated using the LSDA and LDA$+U$ approximations are shown in Fig.~\ref{fig8}. The semi-empirical values corresponding to 7.5\% of the atomic values of the Coulomb-exchange parameter ($U_{eff}$) (see Table~\ref{Tab1}) as reported previously \cite{KFF06} have been used for all given examples. Both compounds exhibit a very small minority band gap at $\epsilon_F$ without $U_{eff}$. However, it becomes large when LDA+$U$ is used in the calculations (for band gaps see Tabs.~\ref{Tab4} and \ref{Tab5}). It should be noted that the high $m$ compounds are very sensitive to electron-electron correlation. In some cases it is required to explain their electronic and magnetic properties. The details of the change in the electronic structure are discussed in Ref.\cite{KFF06}.

In most of the Co$_2$-based Heusler compounds, the minority gap becomes large with inclusion of $U$ (see Tabs.~\ref{Tab4} and \ref{Tab5}). In addition the $\epsilon_F$ may shift to inside or outside the gap. For the particular case of Co$_2$CrAl, $\epsilon_F$ falls outside the gap with $U$.

%%%This tble is not needed, already all results are in previous tables%%%%%%
%\begin{table}
%\centering
%\caption{Minority band gaps (in eV) of Co$_2$FeAl and Co$_2$FeGa with and without $U$.}
%
%     \begin{tabular}{l|cc}
%        compound             & GGA  & LDA$+U$ \\
%        \noalign{\smallskip}\hline\noalign{\smallskip}
%        Co$_2$FeAl        & 0.073 & 0.790 \\
%        Co$_2$MnSi        & 0.798 & 1.304 \\
%        Co$_2$FeGa        & 0.021 & 0.755 \\
%        Co$_2$FeSi        & 0.142 & 0.782 \\
%        \noalign{\smallskip}\hline
%    \end{tabular}
%    \label{tabn}
%\end{table}
%%%%%%%%%%%%%%%%%%%%%%%%%%%%%%%%%%%%%%%%%%%%%%%%%%%
%%%%%%%%%%%%%%%%%%%%%%%%%%%%%%%%%%%%%%%%%%%%%%%%%%%
%%%%%%%%%%%%%%%%%%%%%%%%%%%%%%%%%%%%%%%%%%%%%%%%%%%

\subsection{Other Co$_2$ based half-metallic ferromagnets.}

Half-metallic ferromagnetism may not only being  found in those Co$_2BC$ compounds where $B$ is a $3d$ element but also if $B$ is a $4d$ (Zr, Nb) or $5d$ (Hf) transition metal. There are various Co$_2$-based Heusler compounds of this type reported. Among those, it was found that most of them can be expected to exhibit half-metallic ferromagnetism.

\begin{table}
\caption{Properties of $4d$ or $5d$ containing Co$_2$-based Heusler compounds.
         (For quantities see Tabs.~\ref{Tab2} and \ref{Tab3}.)}
\centering
\begin{tabular}{ l| c | c | c | c | c | c | c | c }
\hline
compound   & $a_{exp}$ & $m_{exp}$ & $m_{calc}$ & $m_A$ & $m_B$  & $N_V$ & $\Delta E$ & Ref.\\
\hline  
Co$_2$ZrAl & 6.078 & 0.79 & 1.00 & 0.62 & -0.10 & 25 & 0.296 & \cite{KSN05}\\
Co$_2$HfAl & 6.045 & 0.82 & 1.00 & 0.61 & -0.09 & 25 & 0.363 & \\
Co$_2$HfGa & 6.032 & 0.60 & 1.00 & 0.60 & -0.09 & 25 & 0.068 & \\
\hline
Co$_2$NbAl & 5.935 & 1.35 & 2.00 & 1.04 &  0.01 & 26 & 0.557 & \cite{MVK83}\\
Co$_2$NbGa & 5.95  & 1.39 & 2.00 & 1.04 & -0.01 & 26 & 0.527 & \\
Co$_2$ZrSn & 6.254 & 1.81 & 2.00 & 1.09 & -0.09 & 26 & 0.540 & \cite{Jei83,SJN99,ZQS06} \\
Co$_2$HfSn & 6.227 & 1.57 & 2.00 & 1.07 & -0.07 & 26 & 0.480 & \cite{GKL74}\\
\hline
Co$_2$NbSn & 6.152 & 0.69 & 1.94 & 0.95 &  0.07 & 27 & 0.430 & \cite{BBF02,AHB05}\\
\hline
\end{tabular}
\label{Tab6}
\end{table}
%%%%%%%%%%%%%%%%%%%%%%%%%%%%%%%%%%%%%%%%%%%%%%%%%%%%%%%%%%%%%%%%%%%%

The magnetic properties of several, reported Co$_2$-based Heusler compounds with $B$ being not a $3d$ transition metal are summarized in Tab.~\ref{Tab6}. There is obviously a large discrepancy between observed and calculated magnetic moments. For Co$_2$NbSn, neither the measured nor the calculated magnetic moment come close to the value expected for a half-metallic state. For the Hf containing compounds, spin-orbit interaction may play already a role, which was not accounted in the calculations. The most interesting point is, however, that only the Co atoms are responsible for the magnetic moment. That means, in none of these compounds a localized magnetic moment at the $B$ atoms is present. This points on the important fact that the existence of a localized moment at the $B$ atoms is not a necessary condition for the occurrence of half-metallic ferromagnetism in Co$_2$-based Heusler compounds.

%%%%%%%%%%%%%%%%%%%%%%%%%%%%%%%%%%%%%%%%%%%%%%%%%%%%%%%%%%%%%%%%%%%%
\subsection{Other Heusler compounds exhibiting half-metallic ferromagnetism}

So far, only the properties of Heusler compounds based on Co$_2$ were considered. However, there exist also half-metallic ferromagnets in the remaining large group of Heusler compounds (Table~\ref{Tab7}). For example, Galanakis {\it et al.} \cite{GDP02} have proposed HMF in Rh$_2$-based compounds.

From the remaining group of known Heusler compounds, half-metallic ferromagnetism is only found in Mn containing compounds. Mn$_2$VAl is the only Heusler compound exhibiting a gap in the majority density of states, unlike the other Heusler compounds. The reason is that Mn$_2$VAl has only 22 valence electrons, that is less than 24, therefore, the completely filled bands appear in the majority states. Indeed, fixing the number of occupied states in the spin channel exhibiting the gap, which has the result that the other channel has to have less electrons occupied. This is expressed in the Slater-Pauling rule where $m$ becomes virtually negative if $N_V<24$.

Mn$_3$Al and Mn$_3$Ga are two binary compounds reported to order in the same space group as Heusler compounds. Both have zero total magnetic moments and thus are half-metallic completely compensated ferrimagnets (HMCCF) \cite{WKF06}. Fe$_2$MnSi and Ir$_2$MnAl are the two non Co$_2$-based Heusler compounds exhibiting HMF behavior. The other family of Heusler compounds studied are the Ru$_2$-based. They all have a gap above the $\epsilon_F$ and are Type~III half metal. Note in this class, Ru atoms carry small spin moments, in some cases the Ru moment is even negative. However, the Mn moment is large and positive and is responsible for the total magnetic moment. There is no compound found from other Ni$_2$, Cu$_2$ and Pd$_2$ based Heusler compounds, which exhibits HMF. Most of them are paramagnetic or ferromagnetic without any gap.

\begin{table}
\caption{Properties of ``other" Heusler compounds exhibiting half-metallic ferromagnetism. The crystal structures of Ru$_2$ based compounds are taken from Ref.~\cite{IKF95}.}
         \centering
\begin{tabular}{ l | c | c | c | c | c | c | c}
\hline
compound   & $a_{cal}$ & $m_{expt}$ & $m_{calc}$ & $m_A$ & $m_B$ &  $\Delta E$ &  $N_V$ \\

\hline
Mn$_2$VAl  & 5.897 & 1.82 & 1.99   & 1.52    & -0.95  & 0.318 & 22 \\
Mn$_2$MnAl & 5.804 &      & 0.002  & 1.423   & -2.836 & 0.546 & 24 \\
Mn$_2$MnGa & 5.823 &      & -0.012 & 1.539   & -3.030 & 0.172 & 24 \\
Mn$_2$MnSi & 5.722 &      & 1.000  & -0.88   & 2.69   & 0.624 & 25 \\
\hline
Fe$_2$MnSi &5.671  & 2.33 & 3.00   & 0.2     & 2.63   & 0.633 & 27 \\
\hline
Ir$_2$MnAl & 6.025 &      & 3.99   & 0.24    & 3.5    & 0.351 & 28 \\
\hline
Ru$_2$MnGe & 5.985 &      & 3.03   & -0.008  & 3.00   & 0.125 & 27 \\
Ru$_2$MnSi & 5.887 &      & 3.00   & 0.02    & 2.92   & 0.097 & 27 \\
Ru$_2$MnSn & 6.217 &      & 3.08   & -0.06   & 3.21   & 0.136 & 27 \\
Ru$_2$MnSb & 6.200 &      & 4.02   & 0.22    & 3.55   & 0.280 & 28 \\
\hline
\end{tabular}
\label{Tab7}
\end{table}
%%%%%%%%%%%%%%%%%%%%%%%%%%%%%%%%%%%%%%%%%%%%%%%%%%%%%%%%%%%%%%%%%%%%

%%%%%%%%%%%%%%% End Section Results and Discussion %%%%%%%%%%%%%%%%%
%%%%%%%%%%%%%%%%%%%%%%%%%%%%%%%%%%%%%%%%%%%%%%%%%%%%%%%%%%%%%%%%%%%%

\section{Conclusions}

In summary, it is proposed that the half-metallic properties in Co$_2BC$ Heusler compounds are dominated by the presence of $C$ atoms. According to the results described here, nearly all Co$_2BC$ compounds will be half-metal ferromagnets.

The magnetic moment carried by $A$ and $B$ atoms is restricted by the $C$ atoms even though they do not directly contribute to the magnetic properties. When the main group element Al from Co$_2B$Al is replaced by Ga, the compound losses its HMF behavior. The influence of $N_V$ on the partial magnetic moments was investigated in detail. It is found that Co$_2BC$ compounds strictly fulfill the Slater-Pauling rule, whereas other compounds exhibit pronounced deviations from the Slater-Pauling type behavior. The minority band gap decreases with increasing lattice parameter. The inclusion of electronic correlation on top of LSDA and GGA, does not destroy the HMF behavior of Co$_2BC$ compounds and non HMF compounds become HMF.

It was found that the existence of a localized moment at the $B$ atoms is not a necessary condition for the occurrence of half-metallic ferromagnetism. Some of these systems with more than 24 valence electrons which exhibit novel magnetic properties, namely half-metallic ferro and ferrimagnetism have been studied in detail. The large exchange splitting of the $B$ atoms are responsible for the half-metallic property of some of these systems.

\begin{acknowledgments}

This work is financially supported by the DFG (project TP7 in research group FG 559).

\end{acknowledgments}

\bibliography{kandpal}% Produces the bibliography via BibTeX.

\end{document}